\documentclass[a4paper,11pt]{article}
\usepackage{geometry}
\usepackage{epsfig}
\usepackage{graphicx}
\usepackage{amsmath}
\usepackage{amssymb}
\usepackage{subfigure}
\usepackage{algorithm} 
\usepackage{algorithmic} 
\usepackage{color}
\usepackage{bm}
\usepackage{color}
\usepackage[hang]{footmisc}

\newtheorem{proof}{Proof}

\title{DCAR: A Discriminative and Compact Audio Representation to Improve Event Detection}

\author{
Liping Jing \thanks{Beijing Key Lab of Traffic Data Analysis and Mining, Beijing Jiaotong University, Beijing, China}
\and
Bo Liu\footnotemark[1]
\and
Jaeyoung Choi  \thanks{International Computer Science Institute, Berkeley, CA, USA and Delft University of Technology, Delft, Netherlands}
\and
Adam Janin \thanks{International Computer Science Institute, Berkeley, CA, USA}
\and
Julia Bernd\footnotemark[3] 
\and
Michael W. Mahoney \thanks{International Computer Science Institute and University of California, Berkeley, CA, USA}
\and
Gerald Friedland \footnotemark[4] 
}

\date{}

\begin{document}
\maketitle

\begin{abstract}

This paper presents a novel two-phase method for audio representation, Discriminative and Compact Audio Representation (DCAR), and evaluates its performance at detecting events in consumer-produced videos. In the first phase of DCAR, each audio track is modeled using a Gaussian mixture model (GMM) that includes several components to capture the variability within that track. The second phase takes into account both global structure and local structure. In this phase, the components are rendered more discriminative and compact by formulating an optimization problem on Grassmannian manifolds, which we found represents the structure of audio effectively. 

Our experiments used the YLI-MED dataset (an open TRECVID-style video corpus based on YFCC100M),
which includes ten events. The results show that the proposed DCAR representation consistently outperforms state-of-the-art audio representations. DCAR's advantage over i-vector, mv-vector, and GMM representations is significant for both easier and harder discrimination tasks. We discuss how these performance differences across easy and hard cases follow from how each type of model leverages (or doesn't leverage) the intrinsic structure of the data. Furthermore, DCAR shows a particularly notable accuracy advantage on events where humans have more difficulty classifying the videos, i.e., events with lower mean annotator confidence.\footnote{An abbreviated version of this paper will be published as Jing et al.\ 2016 \cite{JlLbCjJaBjMmFg}.}

\end{abstract}

\section{Introduction}

With the rapid increase in the number of user-generated videos shared on the Internet, it is becoming increasingly advantageous to explore new ways of retrieving them---for example, by automatically detecting events occurring in them. One less explored approach is to analyze the soundtracks. 
But while the analysis of visual content is widely studied in disciplines such as image processing and computer vision, analysis of video soundtracks 
has largely been restricted to specific, inherently audio-focused tasks 
such as speech processing and music retrieval. 

However, when visual information cannot reliably identify content (e.g., due to poor lighting conditions), audio may still furnish vivid information.
In other words, audio content is frequently complementary to visual content---and in addition, it offers more tractable processing.

Multimodal approaches that use both visual and audio cues have recently gained traction. However, there has not been much in-depth exploration of how best to leverage the audio information.
On the audio side, due to a historical focus on carefully-curated speech and music processing corpora, fewer audio researchers consider the problems posed by unfiltered generic audio with varied background noises---but these problems must be addressed to build audio classifiers that can handle user-generated video. In addition, past work on video analysis has often fed audio ``blindly" into a machine learner, without much consideration of how audio information is structured. 

In the last few years, the focus has been changing in both event detection and audio analysis. For example, sound event detection was included in the 2013 Detection and Classification of Acoustic Scenes and Events (DCASE) challenge at IEEE AASP \cite{SdGdBeLmPm15}. More recently, the YLI-MED annotated video corpus became available \cite{BjBdEbFgGhGlJaKsTjWj15};
YLI-MED is targeted toward multimedia event detection, so provides a good platform to study audio-based event detection (see Section \ref{sec:dataset}).

Major areas of audio-based event detection research include audio data representation and learning methodologies. In this work, we focus on the first aspect, audio data representation, which aims to extract specific features that can refine an enormous amount of raw audio data into higher-level information about the audio signals. Section \ref{sec:related} gives an overview of current representation approaches and discusses their limitations in detail. To offer a brief summary, current approaches do not effectively capture signal variance within audio tracks nor local structure (for example, between Gaussian components), they risk losing information about geometric manifold structure and hidden structure within the data, they often require a lot of storage space, and they rarely leverage available information from labels.

In this paper, we address these issues 
by introducing a Discriminative and Compact Audio Representation (DCAR) to model audio information. This method is implemented in two phases. First, each audio track is modeled using a Gaussian mixture model (GMM) with several mixture components to describe its statistical distribution. This is beneficial for capturing the variability within each audio track and for reducing the storage space required, relative to the full original number of frames. 

Second, by integrating the labels for the audio tracks and the local structure among the Gaussian components, we identify an embedding to reduce the dimensionality of the mixture components and render them more discriminative. In this phase, the dimensionality reduction task is formulated as an optimization problem on a Grassmannian manifold and solved via the conjugate gradient method. 
Then a new audio track can be represented with the aid of the learned embedding, which further compacts the audio information. For classification, we adopt the kernel ridge regression (KRR) method, which is compatible with the manifold structure of the data. 

As we argue in detail in Section~\ref{SecDGMMwhole}, DCAR represents a considerable advancement of the state-of-the-art in audio-based event detection. In a nutshell, the novelty of DCAR lies in its being a \textit{compact representation} of an audio signal that \textit{captures variability} and has \textit{better discriminative ability} than other representations. 

Our claim is supported by a series of experiments, described in Section \ref{SecExp}, conducted on the YLI-MED dataset. 
We first built binary classifiers for each pair of events in the dataset, and found that the proposed DCAR performed better than an i-vector strategy 
on pairwise discrimination.
We then delved deeper, comparing multi-event detection
results for DCAR with three existing methods (including simple GMMs and mean/variance vectors as well as i-vectors)
for events that are difficult to distinguish vs.\ events that are easy to distinguish. 
We showed that DCAR can handle 
both easy and hard cases;
Section~\ref{sec:easyhard} discusses how these results 
may follow from how each type of model leverages (or doesn't leverage) the intrinsic structure of the data. 
Finally, we conducted multi-event detection experiments on all ten events, again showing that DCAR is the most discriminative representation. In particular, DCAR shows notable 
accuracy gains on events where humans find it more difficult to classify the videos, i.e., events with lower average annotator confidence scores.

The remainder of this paper is organized as follows: Section \ref{sec:related} surveys related audio work;  Section \ref{SecDGMMwhole} presents the proposed DCAR model in detail; and Section \ref{sec:EDtask} describes the audio-based event detection process with KRR. Section \ref{sec:datameth} describes our methods and the real-world video dataset YLI-MED; and Section \ref{sec:experiments} discusses a series of binary and multi-event detection experiments. The results demonstrate that DCAR significantly improves event-detection performance. Conclusions and future work are discussed in Section \ref{sec:conc}.

\section{Related Work}
\label{sec:related}

Audio representations include low-level features (e.g., energy, cepstral, and harmonic features)
and intermediate-level features obtained via further processing steps such as filtering, linear combination, unsupervised learning, and matrix factorization (see overview in Barchiesi et al.\ 2015 \cite{BdGdSdPm15}).

A typical audio representation method for event detection is to model each audio file as a vector so that traditional classification methods can be easily applied. 
The most popular low-level feature used is Mel-frequency cepstral coefficients (MFCCs) \cite{EaTjKaFs03}, which describe the local spectral envelope of audio signals. 
However, MFCC is a short-term frame-level representation, so it does not capture the whole structure hidden in each audio signal. 
As one means to address this, some researchers have used end-to-end classification methods (e.g., neural networks), for example to simultaneously learn intermediate-level audio concepts and train an event classifier \cite{RmEbBjFg15}.
Several approaches have used first-order statistics derived from the frames' MFCC features, 
which empirically improves performance on audio-based event detection.
For example, Jin et al.\ adopted a codebook model to define audio concepts \cite{JqSpRsBsDd12}. This method uses first-order statistics to represent audio: it quantizes low-level features into discrete codewords, generated via clustering, and provides a histogram of codeword counts for each audio file (i.e., it uses the mean of the data in each cluster).

However, such methods do not capture the complexity of real-life audio recordings. For event detection, researchers have therefore modeled audio using the second-order statistical covariance matrix of the low-level MFCC features \cite{MrLhGlFgDa11,HzCyLkHvLc13,EbLhFg13,RgNwHp13}. 
There are two ways to compute the second-order statistics.
The first assumes that each audio file can be characterized by the mean and variance of the MFCC features representing each audio frame, then modeled via a vector by concatenating the mean and variance \cite{RgNwHp13}; this representation can be referred to as a mean/variance vector or \textit{mv-vector}. 
The other method is to model all training audio via a Gaussian mixture model and then compute the Baum-Welch statistics of each audio file according to the mixture components, as in GMM-supervector representations \cite{MrLhGlFgDa11}. Again, each audio file is represented by stacking 
the means and covariance matrices. However, such a vectorization process will inevitably distort the geometric structure\footnote{By \textit{geometric structure}, we mean intrinsic structure within data such as affine structure, projective structure, etc.} of the data \cite{SbSa02}. 

An exciting area of recent work is the i-vector approach, which uses latent factor analysis to compensate for foreground and background variability \cite{DnKpDrDpOp11}. The i-vector approach can be seen as an extension of the GMM-supervector. It assumes that these high-dimensional supervectors can be confined to a low-dimensional subspace; this can be implemented by applying  probabilistic principal component analysis (PCA) to the supervectors. 
The advantage of an i-vector is that the system learns the total variance from the training data and then uses it on new data, so that the representation of the new data has similar discriminativity to the representation of the training data. I-vectors have shown promising performance in audio-based event detection \cite{EbLhFg13,HzCyLkHvLc13}.

In fact, many of these representation methods have shown promising performance, but they have some limitations with regard to audio-based event detection.
For example, the signal variance within a given audio track may be large; training a 
Gaussian mixture model on all of the audio tracks (as in the GMM-supervector or i-vector approaches) does not capture that variability, and thus may not characterize a given event well. The second limitation is that each mixture component consists of both a mean vector and a covariance matrix, which can introduce many more variables, and so result in high computational complexity and require a lot of storage space. The third limitation is that the covariance matrices of the mixture components in these methods are usually flattened into one supervector, which may distort the geometric manifold structure within the data and lose information about hidden structure.
Fourth, most audio representations are derived in an unsupervised manner,
i.e., they do not make use of any existing label information. But in fact, label information has been very useful for representing data in classification tasks such as image classification \cite{JlZcMn12} and text classification \cite{LmTcSjLy09}. Last but not least, these methods do not explicitly consider the local structure between Gaussian components, which may be useful for distinguishing events.

These drawbacks motivate us to propose a new audio representation method to capture the variability within each audio file and to characterize the distinct structures of events with the aid of valuable existing labels and local structure within the data; these characteristics of our method have significant benefits for event detection.

\section{Discriminative and Compact Audio Representation}\label{SecDGMMwhole}

In this section, we describe our proposed two-phase audio representation method. The first phase, described in Subsection \ref{SecGMM}, aims to capture the variability within each audio file. The second phase, described in Subsection \ref{SecDGMM}, identifies a discriminative embedding. 

\subsection{Phase 1: Characterizing Per-Track Variability}
\label{SecGMM}

Given a set of audio tracks, we first extract their low-level features, in this case MFCC features. Let $\mathbf{X}=\{\mathbf{X}^i\}_{i=1}^n$ denote a set of $n$ audio files. Each file $\mathbf{X}^i$ is segmented into $m_i$ frames. Each frame is computed using a 100ms Hamming window with a stride size of 10 ms per frame shift, and its corresponding representation $x^i_j$ ($1\leq i \leq n$ and $1 \leq j \leq m_i$) is built with the first 20 MFCC features and their first-order and second-order derivatives. Each frame is then modeled via a vector with $d$-dimensional MFCC features ($d=60$), i.e., $x^i_j \in \mathbb{R}^{60}$.

Previous work has demonstrated that second-order statistics are much more appropriate for describing complicated multimedia data \cite{DnKpDrDpOp11,WwWrHzSsCx15}. Therefore, we train a GMM with $P$ components using the Expectation-Maximization algorithm for each audio track:
\begin{equation}\label{eq:ExMax}
\mathbf{X}^i = \{x_j^i\}_{j=1}^{m_i}\in \mathbb{R}^{d\times m_i}.
\end{equation}
The estimated GMM components are denoted as: 
\begin{equation}\label{eq:GMMcomps}
G=\{g_i\}_{i=1}^N
\end{equation}
where $g_i=\{w_i,\mu_i,\Sigma_i\}$. When each audio file is modeled via $P$ components, $N=nP$. Each component has its corresponding weight $w_i$, mean $\mu_i$, and covariance matrix $\Sigma_i$. 

Generally, covariance matrices are positive semi-definite, and can be made strictly positive definite by adding a small constant to the diagonal elements of the matrix. 
For convenience, we use the notation $\Sigma$ to indicate a symmetric positive definite (SPD) matrix.
After GMM modeling, each audio file---typically containing hundreds to thousands of frames---is reduced to a smaller number of mixture components with prior probabilities. The covariance matrices provide a compact and informative feature descriptor, which lies on a specific manifold, and obviously captures the (second-order) variability of the audio.

\subsection{Phase 2: Identifying a Discriminative Embedding}\label{SecDGMM}

In the second phase of the DCAR method, a discriminative embedding is identified by integrating the global and local structure of the training data, so that both training data and unseen new data can be re-represented in a discriminative and compact manner.
 
\subsubsection{Overview of Phase 2}
\label{sec:DCARovw}

\begin{figure}[!htb] 
   \centering
   \includegraphics[width=0.4\textwidth]{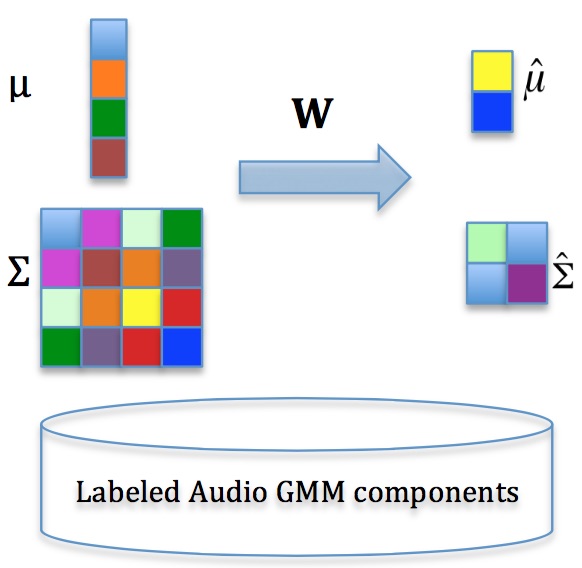}
    \caption{Framework for generating the discriminative and compact audio representation (DCAR). The left side shows the original $d$-dimensional GMM components ($\mu\in \mathbb{R}^d$ and $\Sigma\in \mathbb{R}^{d\times d}$); the right side shows the DCAR representation with $r$-dimensional ($r<d$) mixture components ($\hat{\mu}\in \mathbb{R}^r$ and $\hat{\Sigma}\in \mathbb{R}^{r\times r}$).} \label{Fig:DLGMM}
\end{figure}

Although an audio file can be represented via the above mixture components, the model presented thus far ignores the global structure of the data (e.g., the valuable label information) and the local structure among the components (e.g., nearest neighbors). Meanwhile, the original feature representation is usually large (since there are 60 MFCC features, each mean vector has $60$ elements, and each covariance matrix contains $60\times 60$ elements), which may be time-consuming in later data processing. Therefore, in this subsection, we propose a new method for generating a discriminative and compact representation from the high-dimensional mixture components.
The DCAR method is summarized in Figure \ref{Fig:DLGMM}. 

Our main goal is to learn an embedding $\mathbf{W}\in\mathbb{R}^{d\times r}$ ($r<d$, where $d$ is the number of MFCC features and $r$ is the embedding space size) based on the GMM components of the labeled audio track ($G=\{g_i,\ell_i\}_{i=1}^N$) (where
$\ell_i$ is the label for component $g_i$, based on the label of the corresponding audio file from which $g_i$ was generated). Therefore, the resulting low-dimensional GMM components should preserve the important structure of the original GMM components as much as possible. To accomplish this, we introduce an embedding $\mathbf{W}$ and define the new GMM components with mean
\begin{equation}\label{meanMap}
\hat{\mu} = \mathbf{W}^T\mu
\end{equation}
and covariance matrix
\begin{equation}\label{SigmaMap}
\hat{\Sigma} = \mathbf{W}^T\Sigma \mathbf{W}.
\end{equation}
As mentioned above, the covariance matrix $\Sigma$ is SPD, i.e, $0\prec \Sigma \in \mathcal{S}ym_{d}^+$. To maintain this property, i.e., $0\prec \hat{\Sigma} \in \mathcal{S}ym_{r}^+$, the embedding $\mathbf{W}$ is constrained to be full rank. A simple way of enforcing this requirement is to impose orthonormality constraints on $\mathbf{W}$ (i.e., $\mathbf{W}^T\mathbf{W}=\mathbf{I}_r$), so that the embedding can be identified by solving an optimization problem on the Grassmannian manifold.

For event detection, each training file has label information, which we also assign to its GMM components. This valuable information can be interpreted as global structure for those components. There is also intrinsic internal structure among the components, such as the affinity between each pair of components. When reducing the dimensionality of GMM components, it is necessary to maintain these two types of structure. Motivated by the idea of linear discriminative analysis \cite{McLachlanG2004} and Maximum Margin Criterion \cite{LhJtZk04}, DCAR aims to minimize the intra-class distance while simultaneously maximizing the inter-class distance. In the next subsection, we introduce an undirected graph defined by a real symmetric affinity matrix $\mathbf{A}\in \mathbb{R}^{N\times N}$ that we use to encode these structures.

\subsubsection{Affinity Matrix Construction}\label{SecAMC}

The affinity matrix $\mathbf{A}$ is defined by building an intra (within)-class similarity graph and an inter (between)-class similarity graph, as follows. 
\begin{equation}\label{AffinityGraph}
\mathbf{A}_{ij} = \mathbf{S}_w - \mathbf{S}_b
\end{equation}
$\mathbf{S}_w$ and $\mathbf{S}_b$ are two binary matrices describing the intra-class and inter-class similarity graphs respectively, formulated as:
\begin{equation}
\mathbf{S}_w(g_i,g_j)=  \begin{cases}
    1     & \text{ if } g_i \in \text{NN}_w(g_j) \text{ or } g_j \in \text{NN}_w(g_i) \\
    0     & \text{ otherwise}
  \end{cases}
\end{equation}
and
\begin{equation}
\mathbf{S}_b(g_i,g_j)=  \begin{cases}
    1     & \text{ if } g_i \in \text{NN}_b(g_j) \text{ or } g_j \in \text{NN}_b(g_i) \\
    0     & \text{ otherwise,}\\
  \end{cases}
\end{equation}
where $\text{NN}_w(g_i)$ contains the $n_w$ nearest neighbors of component $g_i$ that share the same label as $\ell_i$, and $\text{NN}_b(g_i)$ is the set of $n_b$ nearest neighbors of $g_i$ that have different labels. Here, the nearest neighbors of each component can be identified via their similarity. We use heat kernel weight with a self-tuning technique (for parameters $\sigma_{\mu}$ and $\sigma_{\Sigma}$) to measure the similarity between components:
\begin{equation}\label{GaussianKernel}
S(g_{i},g_{j}) = \lambda \exp\Big (\frac{-\delta_{\mu}^2(\mu_{i},\mu_{j})}{2\sigma_{\mu}^2} \Big) + \exp\Big (\frac{-\delta_{\Sigma}^2(\Sigma_{i},\Sigma_{j})}{2\sigma_{\Sigma}^2} \Big)
\end{equation}
where $\lambda$ is a trade-off parameter to control the contribution from the components' means and covariance matrices and $\delta_{\mu}$ indicates the distance measure for the means of the mixture components. Here we use the simple euclidean distance
\begin{equation}\label{DisMu}
\delta_{\mu}^2(\mu_i,\mu_j)=\| \mu_i - \mu_j\|_2^2.
\end{equation}
$\delta_{\Sigma}$ indicates the distance measure for the covariance matrices of the components. 

A number of metrics have been used in previous research, including the Affine-Invariant Riemannian Metric (AIRM) \cite{PxFpAn06}, Stein Divergence \cite{CaSsBaPn11}, and the Log-Euclidean Metric (LEM) \cite{AvFpPxAn07}. AIRM imposes a high computational burden in practice, and we have observed experimentally that nearest neighbors selected according to LEM more often fall into the same event than nearest neighbors selected according to either AIRM or Stein (see Appendix \ref{app:comparison} for details). 
For these two reasons, we exploit LEM to compute $\delta_{\Sigma}$:
\begin{equation}\label{DisSigma}
\delta_{\Sigma}^2(\Sigma_i,\Sigma_j)=\| \log(\Sigma_i)-\log(\Sigma_j)\|_F^2.
\end{equation}

The constructed affinity matrix $\mathbf{A}$ thus effectively combines local structure, i.e., nearest neighbors, and global structure, i.e., label information---which is used to find the within-class nearest neighbor ($\text{NN}_w$) and the between-class nearest neighbor ($\text{NN}_b$).

\subsubsection{Embedding Optimization}

Once we have $\mathbf{A}$, the next step is to learn an embedding such that the structure among the original GMM components $\{g_i\}_{i=1}^N = \{\mu_i,\Sigma_i\}_{i=1}^N$ is reflected by the low-dimensional mixture components $\{\hat{g}_i\}_{i=1}^N = \{\hat{\mu}_i,\hat{\Sigma}_i\}_{i=1}^N$. This process can be modeled using the following optimization problem:
\begin{equation}\label{DLDGMM1}
\mathbf{F}(\mathbf{W}) = \min\limits_{\mathbf{W}^T\mathbf{W}=I_{r}}\sum_{i,j}A_{ij}\Big (\lambda \delta_{\mu}^2(\hat{\mu}_i,\hat{\mu}_j) +  \delta_{\Sigma}^2(\hat{\Sigma}_i,\hat{\Sigma_j})\Big )
\end{equation}
With the aid of the mapping functions in (\ref{meanMap}) and (\ref{SigmaMap}) and the distance metrics $\delta_{\mu}$ (\ref{DisMu}) and $\delta_{\Sigma}$ (\ref{DisSigma}), the optimization problem can be rewritten as:
\begin{equation}\label{DLDGMM2}
\begin{split}
\mathbf{F}(\mathbf{W}) = 
& \min\limits_{\mathbf{W}^T\mathbf{W}=I_{r}}\sum_{i,j}A_{ij}\Big (\lambda \|\mathbf{W}^T(\mu_i-\mu_j)\|_F^2 \\
& + \|\log(\mathbf{W}^T\Sigma_i\mathbf{W}) - \log(\mathbf{W}^T\Sigma_j\mathbf{W})\|_F^2\Big )
\end{split}
\end{equation}
As in (\ref{GaussianKernel}), $\lambda$ is used to balance the effects of two terms, in tuning by cross-validation on the training data.
Optimizing $\mathbf{F}(\mathbf{W})$ results in a situation where the low-dimensional components are close if their corresponding original high-dimensional components are event-aware neighbors; otherwise, they will be as far apart as possible.

In image processing, there are several lines of research where a mapping has been learned from a high-dimensional manifold to a low-dimensional manifold \cite{HmSmHr14,HzWrSsLxCx15}. However, Harandi et al.\ \cite{HmSmHr14} exploit AIRM and Stein Divergence to measure the distance, and as we noted in Subsection \ref{SecAMC}, these metrics are not appropriate for handling audio data. Huang et al.\ \cite{HzWrSsLxCx15} identified an embedding on the logarithms of the SPD matrix, but our goal is to identify an embedding on GMM components, including both means and covariance matrices.

The problem in (\ref{DLDGMM2}) is a typical optimization problem with orthogonality constraints; it can
therefore be formulated as a unconstrained optimization problem on Grassmannian manifolds \cite{ApMrSr08}. Given that the objective function $\mathbf{F}(\mathbf{W})$ has the property that for any rotation matrix $\mathbf{R}\in SO(r)$  (i.e., $\mathbf{R}\mathbf{R}^T=\mathbf{R}^T\mathbf{R}=\mathbf{I}_r$), $\mathbf{F}(\mathbf{W})=\mathbf{F}(\mathbf{WR})$ (see Appendix \ref{app:invariance} for a detailed proof),
this optimization problem is most compatible with a Grassmannian manifold.
In other words, we can model the embedding $\mathbf{W}$ as a point on a Grassmannian manifold $\mathcal{G}(r,d)$, which consists of the set of all linear $r$-dimensional subspaces of $\mathbb{R}^d$. 

Here we employ the conjugate gradient (CG) technique to solve (\ref{DLDGMM2}), because CG is easy to implement, has low storage requirements, and provides superlunar convergence in the limit \cite{ApMrSr08}. On a Grassmannian manifold, the CG method performs minimization along geodesics with specific search directions. Here, the geodesic is the shortest path between two points on the manifold. For every point on the manifold $\mathcal{G}$, its tangent space is a vector space that contains the tangent vectors of all possible curves passing through that point. Unlike flat spaces, on a manifold we cannot directly transport a tangent vector from one point to another point by simple translation; therefore, the tangent vectors must parallel transport along the geodesics. 

More specifically, on the Grassmannian manifold, let $\nabla_{\mathbf{W}}$ and $\mathcal{D}_{\mathbf{W}}$ be the tangent vector and the gradient of $\mathbf{F}(\mathbf{W})$ at point $\mathbf{W}$, respectively. The gradient on the manifold at the $\tau$-th iteration can be obtained by subtracting the normal component at $\mathbf{W}^{(\tau)}$ from the transported vector: 
\begin{equation}\label{GradientWonM}
\mathcal{D}_{\mathbf{W}}^{(\tau)} = \nabla_{\mathbf{W}}^{(\tau)} - \mathbf{W}^{(\tau)}(\mathbf{W}^{(\tau)})^T\nabla_{\mathbf{W}}^{(\tau)}.
\end{equation}
Then the search direction $\mathcal{H}_{\mathbf{W}}$ in the $(\tau+1)$-th iteration can be computed by parallel transporting the previous search direction and combining it with the gradient direction at the current solution:  
\begin{equation}\label{SearchDirectionWonM}
\mathcal{H}_{\mathbf{W}}^{(\tau+1)} =  \mathcal{D}_{\mathbf{W}}^{(\tau+1)} + \gamma^{(\tau+1)}\triangle\mathcal{H}_{\mathbf{W}}^{(\tau)}.
\end{equation}
Here $\triangle\mathcal{H}_{\mathbf{W}}^{(\tau)}$ is the parallel translation of the vector $\mathcal{H}_{\mathbf{W}}^{(\tau)}$. According to Absil et al.\ \cite{ApMrSr08}, the geodesic going from point $\mathbf{W}$ in the direction $\mathcal{H}_{\mathbf{W}}^{(\tau)}$ can be represented by the geodesic equation
\begin{equation}\label{WangleT}
\mathbf{W}(t) = \left [  \begin{array}{cc}
\mathbf{W}V &U \end{array}\right ] \left [  \begin{array}{c}
\cos \Lambda t \\ \sin \Lambda t\end{array}\right ] V^T.
\end{equation}
Thus, the parallel translation can be obtained by
\begin{equation}\label{SearchDirectionParTraWonM}
\triangle\mathcal{H}_{\mathbf{W}}^{(\tau)} = \big( -\mathbf{W}^{(\tau)} V \sin \Lambda t^{(\tau)} + U \cos \Lambda t^{(\tau)} \big)\Lambda V^T
\end{equation}
where $U\Lambda V^T$ is the compact singular value decomposition of $\mathcal{H}_{\mathbf{W}}^{(\tau)}$.

We use the exact conjugacy condition to adaptively determine the step size $\gamma^{(\tau+1)}$, as follows:
\begin{equation}\label{StepSizeWonM}
\gamma^{(\tau+1)} = \frac{< \mathcal{D}_{\mathbf{W}}^{(\tau+1)}- \triangle\mathcal{D}_{\mathbf{W}}^{(\tau)},  \mathcal{D}_{\mathbf{W}}^{(\tau+1)}>}{< \mathcal{D}_{\mathbf{W}}^{(\tau)}, \mathcal{D}_{\mathbf{W}}^{(\tau)}>}
\end{equation}
where $<A,B>=Tr(A^TB)$. 
Similar to $\triangle\mathcal{H}_{\mathbf{W}}^{(\tau)}$, $\triangle\mathcal{D}_{\mathbf{W}}^{(\tau)}$ is the parallel translation of the vector $\mathcal{D}_{\mathbf{W}}^{(\tau)}$ on the Grassmannian manifold, which can be calculated thusly:
\begin{equation}\label{TangetVectParTraWonM}
\begin{small}
\triangle\mathcal{D}_{\mathbf{W}}^{(\tau)} = \mathcal{D}_{\mathbf{W}}^{(\tau)} - \big( \mathbf{W}^{(\tau)} V \sin \Lambda t^{(\tau)} + U (\mathbf{I}-\cos \Lambda t^{(\tau)}) \big)U^T \mathcal{D}_{\mathbf{W}}^{(\tau)} 
\end{small}
\end{equation}

Going back to the objective function in (\ref{DLDGMM2}), by setting $\mathbf{F1}_{ij} = \|\mathbf{W}^T(\mu_i-\mu_j)\|_F^2$ and $\mathbf{F2}_{ij} = \|\log(\mathbf{W}^T\Sigma_i\mathbf{W}) - \log(\mathbf{W}^T\Sigma_j\mathbf{W})\|_F^2$, (\ref{DLDGMM2}) can be rewritten as
\begin{equation}\label{eq:reopt}
\mathbf{F}(\mathbf{W}) =\min\limits_{\mathbf{W}^T\mathbf{W}=I_{m}}\sum_{i,j}A_{ij}\big (\lambda \mathbf{F1}_{ij} +  \mathbf{F2}_{ij} \big ).
\end{equation}
Then its tangent vector $\nabla_{\mathbf{W}}$ on the manifold can be computed in three steps (see Appendix \ref{app:F2} for details):
\begin{equation}\label{DerivJ1}
\nabla_{\mathbf{W}}\mathbf{F1}_{ij} = 2 (\mu_i-\mu_j)(\mu_i-\mu_j)^T\mathbf{W}
\end{equation}
\begin{equation}\label{DerivJ2}
	\begin{split}
	\nabla_{\mathbf{W}}\mathbf{F2}_{ij}
         =4\Big (\Sigma_i\mathbf{W}(\mathbf{W}^T\Sigma_i\mathbf{W})^{-1}-\Sigma_j\mathbf{W}(\mathbf{W}^T\Sigma_j\mathbf{W})^{-1}\Big ) 
	\\  \times \Big(\log(\mathbf{W}^T\Sigma_i\mathbf{W})-\log(\mathbf{W}^T\Sigma_j\mathbf{W})\Big )	\end{split}
\end{equation}
\begin{equation}\label{DerivF}
\nabla_{\mathbf{W}}= \sum_{i,j}A_{ij}\big (\lambda_1 \nabla_{\mathbf{W}}\mathbf{F1}_{ij} + \lambda_2 \nabla_{\mathbf{W}}\mathbf{F2}_{ij} \big )
\end{equation}

This conjugate gradient method for solving (\ref{DLDGMM2}) is summarized in Algorithm \ref{AlgorithmDGMM}.

\begin{algorithm}[h]
	\caption{Solving (\ref{DLDGMM2}) via a Conjugate Gradient on a Grassmannian Manifold}  \label{AlgorithmDGMM}
		\renewcommand{\algorithmicrequire}{\textbf{Input:}}
		\renewcommand\algorithmicensure {\textbf{Output:}}
		\begin{algorithmic}[1]
			\REQUIRE
			A set of labeled $d$-dimensional GMM components $\{g_i,\ell_i\}_{i=1}^N$ with means $\{\mu_i\}_{i=1}^N$ and covariance matrices $\{\Sigma\}_{i=1}^N$, reduced dimensionality $r$, and parameter $\lambda$.
       			\STATE Construct an affinity matrix $A$ using (\ref{AffinityGraph})
			
			\STATE Initialize $\bm W^{(0)}$ such that $(W^{(0)})^TW^{(0)}=\mathbf{I}_r$ and $\tau=0$
			
			\STATE Compute $\nabla_{\mathbf{W}}^{(0)}$ as in (\ref{DerivF}) and $\mathcal{D}_{\mathbf{W}}^{(0)}$ as in (\ref{GradientWonM}), and set  $\mathcal{H}_{\mathbf{W}}^{(0)} = -  \mathcal{D}_{\mathbf{W}}^{(0)}$ 
			
			\FOR{$\tau=0,1,2,\cdots$}
			\STATE Perform compact singular value decomposition of $\mathcal{H}_{\mathbf{W}}^{(\tau)}$
			\STATE Identify $\mathbf{W}^{(\tau)}$  by minimizing (\ref{DLDGMM2}) over $t$, where $\mathbf{W}^{(\tau)}(t)$ is computed as in (\ref{WangleT}), letting $t^{(\tau)} = t_{min}$ (here $t_{min}$ is the value of $t$ minimizing (\ref{DLDGMM2})) 
			\STATE Compute the parallel translations of $\mathcal{H}_{\mathbf{W}}^{(\tau)}$ and $\mathcal{D}_{\mathbf{W}}^{(\tau)}$, i.e., compute $\triangle\mathcal{H}_{\mathbf{W}}^{(\tau)}$ as in (\ref{SearchDirectionParTraWonM}) and $\triangle\mathcal{D}_{\mathbf{W}}^{(\tau)}$ as in (\ref{TangetVectParTraWonM})
			\STATE Set $\mathbf{W}^{(\tau+1)} = \mathbf{W}^{(\tau)} (t^{(\tau)})$
			\STATE Compute $\nabla_{\mathbf{W}}^{(\tau+1)}$ as in (\ref{DerivF}) and $\mathcal{D}_{\mathbf{W}}^{(\tau+1)}$ as in (\ref{GradientWonM})
			\STATE Find the step size $\gamma^{(\tau+1)}$ via (\ref{StepSizeWonM}) 
			\STATE Find the new search direction $\mathcal{H}_{\mathbf{W}}^{(\tau+1)}$ via (\ref{SearchDirectionWonM})
			
			\ENDFOR
			\ENSURE $\bm W^{(\tau+1)}$
		\end{algorithmic}
	\end{algorithm}

Then, given a new audio file, we can extract its MFCC features, train $P$ GMM components, and re-represent these components with the embedding $\bm W$ to get its discriminative, low-dimensional mixture components, i.e., the proposed DCAR representation.

\section{Event Detection with DCARs}
\label{sec:EDtask}

As we describe above, each audio file is represented via several mixture components, including mean vectors and covariance matrices. It would be possible to flatten the matrices into vectors and then use traditional vector-based classification methods for event detection. However, the covariance matrices lie on the manifold of positive definite matrices, and such a vectorization process would ignore this manifold structure \cite{SbSa02}.  
Therefore, we use the Kernel Ridge Regression (KRR) method to build the event classifiers. 

Let $\hat{G}=\{\hat{g}_i\}_{i=1}^N$ and $\hat{g}_i= \{\hat{\mu}_i,\hat{\Sigma}_i\}$ be mixture components for the training audio tracks belonging to $L$ events. 
$Y\in \mathbb{R}^{N\times L}$ indicates the label information for $\hat{G}$, where $\mathbf{Y}_{ij}=1$ if the $i^{\text{th}}$ component belongs to the $j^{\text{th}}$ event; otherwise $\mathbf{Y}_{ij}=0$. The KRR method aims to train a classifier by solving the following optimization problem:
\begin{equation}\label{KernelDA}
\min_{\mathbf{H}}\mathbf{J}(\mathbf{H}) =\|\phi(\hat{G})^T \mathbf{H}-\mathbf{Y}\|_F^2 + \alpha \|\mathbf{H}\|_F^2
\end{equation}
where $\phi$ is a feature mapping from the original feature space to a 'ensional space, and the kernel function can be written as $K=\phi(\hat{G})^T\phi(\hat{G})$. 

Since each component $\hat{g}_i$ has a mean $\hat{\mu}_i$ and a covariance matrix $\hat{\Sigma}_i$, we can define a combined kernel function to integrate these two parts, as follows:
\begin{equation}\label{KernelIntegrated}
K(\hat{g}_{i},\hat{g}_{j}) =\lambda K_{\mu}(\hat{\mu}_{i},\hat{\mu}_{j})  + K_{\Sigma}(\hat{\Sigma}_{i},\hat{\Sigma}_{j})
\end{equation}
The trade-off parameter $\lambda$ (see (\ref{GaussianKernel}) can be tuned by cross-validation on the training data. As described in Section \ref{SecAMC}, we use a Gaussian kernel to calculate $K_{\mu}$ and $K_{\Sigma}$ via
\begin{equation}\label{eq:Kmu}
K_{\mu}(\hat{\mu}_{i},\hat{\mu}_{j}) = \exp\Big (\frac{-\| \hat{\mu}_i - \hat{\mu}_j\|_2^2}{2\sigma_{\hat{\mu}}^2} \Big)
\end{equation}
and
\begin{equation}\label{eq:KSigma}
K_{\Sigma}(\hat{\Sigma}_{i},\hat{\Sigma}_{j}) = \exp\Big (\frac{-\| \log(\hat{\Sigma}_i)-\log(\hat{\Sigma}_j)\|_F^2}{2\sigma_{\hat{\Sigma}}^2} \Big).
\end{equation}
The problem in (\ref{KernelDA}), as a quadratic convex problem, can be optimized by setting its derivative with respect to $H$ to zero, and then computing $H$ in closed form:
\begin{equation}\label{eq:Hcomp}
\mathbf{H} = \phi(\hat{G})(K+\alpha \mathbf{I})^{-1}\mathbf{Y}
\end{equation}
Here $K_{ij} = K(\hat{g}_{i},\hat{g}_{j})$ as given in (\ref{KernelIntegrated}). 

Given a new test audio track, $P$ mixture components $\{g_p\}_{p=1}^P=\{w_p,\mu_p,\Sigma_p\}_{p=1}^P$ can be obtained via the methods described in Section \ref{SecGMM}. Then the corresponding discriminative, low-dimensional mixture components $\{\hat{g}_p\}_{p=1}^P$ can be generated as in (\ref{meanMap}) for $\hat{\mu}_p=W^T\mu_p$, and as in (\ref{SigmaMap}) for $\hat{\Sigma}_p=W^T\Sigma_pW$, where the embedding $\mathbf{W}$ is learned from the training data. 
Then the class membership matrix $M = \{M_p\}_{p=1}^P$ (where $M_p \in \mathbb{R}^{1\times L}$ is the event membership of the $p$-th component) can be calculated:
\begin{equation}\label{eq:MbrMatrix}
M_p = \phi(\hat{g}_p)^T\mathbf{H} = \phi(\hat{g}_p)^T\phi(\hat{G})(K+\alpha \mathbf{I})^{-1}\mathbf{Y} = K_p(K+\alpha \mathbf{I})^{-1}\mathbf{Y}.
\end{equation}
Here $K_p = [K(\hat{g}_{p},\hat{g}_{i})]_{i=1}^N$, indicating the similarity between $\hat{g}_p$ and all of the training mixture components in $\hat{G}$.
We can then make a final prediction about the event of the new audio track with $P$ components using an average voting scheme
\begin{equation}\label{eq:AvgVote}
\ell = arg\max_j\sum_{p=1}^Pw_pM_p(j)
\end{equation}
where $w_p$ is the weight of the $p^{\text{th}}$ component.

\section{Data and Experimental Methods}\label{sec:datameth}

We evaluated the event detection performance of our proposed representation against several baseline representations, using the recently released public data set YLI-MED.

\subsection{Dataset}
\label{sec:dataset}

YLI-MED \cite{BjBdEbFgGhGlJaKsTjWj15} is an open video corpus for multimedia event detection research (modeled on the TRECVID MED corpus \cite{OpAgFjAbMmSaKwQg12}, but publicly available); the videos in it are drawn from the YFCC100M dataset \cite{TbSdFgEbNkPdBdLl16}. YLI-MED includes about 2000 videos that contain examples of ten events, with standard training and test videos for each event. Since this work focuses on analyzing the acoustic environment of videos, we conducted a series of experiments using the audio tracks. Table \ref{dataS} describes the data we used, including the number of training and test audio files and the range in length for the training and testing sets for each event. The wide variation in length among the tracks makes the event detection task more challenging.

\begin{table*}[!thb]
\centering
\small
\caption{Dataset Composition} \label{dataS}
\begin{tabular}{c|l|cc|cc}
\hline
      Event         &                       &  \multicolumn{2}{c|}{Training  Data}  &  \multicolumn{2}{c}{Testing  Data}\\\cline{3-6}
ID  &Event Name  & \# of Videos & length (ms) & \# of Videos  & length (ms) \\\hline
Ev101     & Birthday Party & 99 &6850$\sim$248950 &131 &8380$\sim$328960\\
Ev102     & Flash Mob       & 91	&8290$\sim$325630 &49   &11710$\sim$152560\\
Ev103     & Getting a Vehicle Unstuck & 89 &5590$\sim$591670 &39 &11170$\sim$157690\\
Ev104     & Parade            &95	&7840$\sim$303850 &127 &5770$\sim$216460\\
Ev105     & Person Attempting a Board Trick &99  &5950$\sim$391150  &88 &5500$\sim$254980\\
Ev106     & Person Grooming an Animal        &97  &5950$\sim$574300  &38 &7210$\sim$292870\\
Ev107     & Person Hand-Feeding an Animal  &95 &6850$\sim$174880  &113 &7840$\sim$244450\\
Ev108     & Person Landing a Fish &99 &7930$\sim$363610 &41 &7480$\sim$250120\\
Ev109     & Wedding Ceremony     &90 &9640$\sim$631630 &108  &9820$\sim$646300\\
Ev110     & Working on a Woodworking Project &98 &5590$\sim$373690 &44 &6760$\sim$281080\\
\hline
\end{tabular}
\end{table*}

\subsection{Methodology}
\label{sec:expmethods}

To evaluate our proposed DCAR method, we compared it with three state-of-the-art audio representations used for event detection: mv-vector \cite{RgNwHp13}, i-vector \cite{EbLhFg13}, 
and GMM. By \textit{GMM}, we mean here the base GMMs obtained by extracting the GMM components from each audio file (as described in Section \ref{SecGMM}), but without the discriminative dimensional reduction step used in DCAR (described in Section \ref{SecDGMM}). 

As we mentioned in Section \ref{sec:related}, an i-vector models all of the training audio frames to obtain a GMM supervector for each frame, then factorizes them to obtain the i-vector representation. In contrast, an mv-vector models each audio file via the mean and variance of the MFCC features, then concatenates the mean and variance to obtain a vector representation.\footnote{As we want to evaluate the most comparable aspects of the representations, we do not consider the temporal information from the RQA features for the mv-vector method.} 

There are several parameters for each of the representations, which we tuned using cross-validation on the training data to obtain the best result. For GMM and DCAR, the number of components 
for each audio track is tuned from 1 to 10, with a step of 1. For i-vector, the number of components
in all of the training data is tuned to one of the values in
$\{2^7,2^8,2^9,2^{10}\}$ and the vector dimensionality is tuned to one of the values in
$\{200,400,600,800,1000\}$. For DCAR, the number of nearest neighbors ($n_w$ and $n_b$) is set to be 5 for affinity matrix construction, the embedding space size $r$ is tuned in $[L,60]$ with a step of 5 ($L$ is the number of events), and the trade-off parameter $\lambda$ is tuned to one of $\{10^k\}_{k=-2}^2$.\footnote{In addition, we tried normalizing each track, with each MFCC feature having a mean of 0 and a variance of 1. All GMM components then have 0 means, so the KRR classifier depends solely on the covariance matrix. However, for event detection, information about means appears to be important. When we tuned the trade-off parameter ($\lambda$ in Eq. (\ref{KernelIntegrated})), we found that we usually obtained the best results with both means and covariance matrices.}

Because our focus is on comparing different audio representations, we describe here experiments that all used the same classification method, KRR.\footnote{To check the validity of this approach, we also tested several other classification techniques with mv-vector and i-vector representations, including SVM (support vector machines), KNN (k--nearest neighbor), and PLDA (parallel latent Dirichlet allocation).
The performance rankings between representations were parallel across all classification techniques.}
For the $l$-th event in $t$ testing tracks, we compared the prediction result to the ground truth to determine the number of true positives ($TP_l$), false positives ($FP_l$), true negatives ($TN_l$), and false negatives ($FN_l$).  
We then evaluated event detection performance using four metrics, $Accuracy$, $FScore$, False Alarm Rate ($FAR$), and $MissRate$, defined (respectively) as:
\begin{equation}\label{eq:AccuracyCalc}
Accuracy = \frac{\sum_{l=1}^L TP_l}{t}
\end{equation}
\begin{equation}\label{eq:FScoreCalc}
FScore_l = \frac{2\times TP_l}{2\times TP_l+FP_l+FN_l}
\end{equation}
\begin{equation}\label{eq:FARCalc}
\quad FAR_l = \frac{FP_l}{TN_l+FP_l}
\end{equation}
\begin{equation}\label{eq:MissRateCalc}
\quad MissRate_l= \frac{FN_l}{FN_l+TP_l}.
\end{equation}
$Accuracy$ is calculated on all $t$ testing tracks (i.e., we use combined or overall accuracy), and the other three metrics are calculated for each event and then averaged over $L$ events to evaluate performance. 
Larger $Accuracy$ and $FScore$ values indicate better performance, and smaller $FAR$ and $MissRate$ values indicate better performance. 

\begin{figure*}[!htb]
   \centering
    \subfigure[I-vector (lower triangle) vs.\ DCAR (upper triangle)]{
        \includegraphics[height=0.37\textwidth, width=0.48\textwidth]{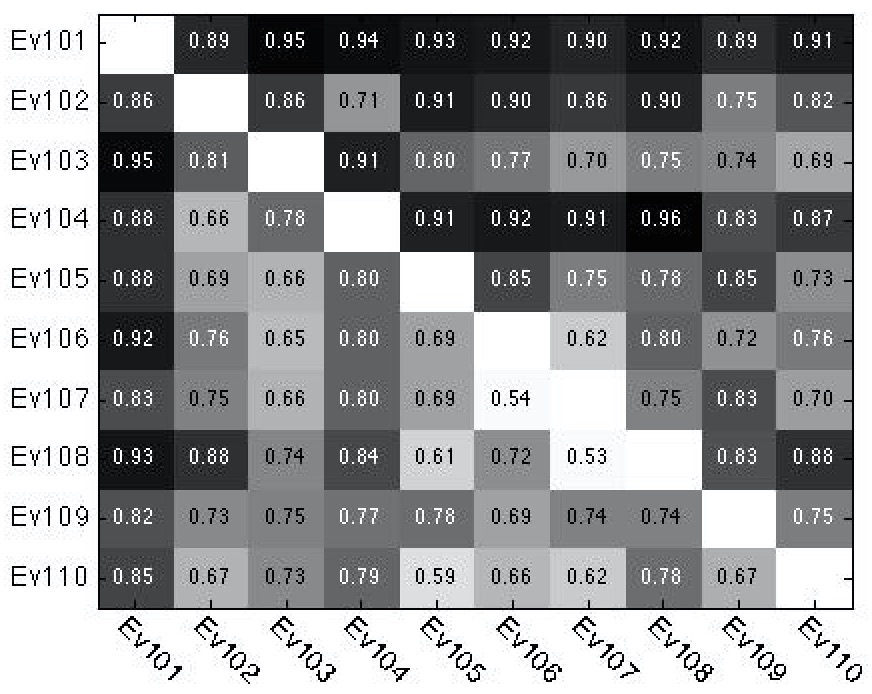}}	
   \subfigure[Histogram of Relative Gain (DCAR vs.\ i-vector)]{
	\includegraphics[height=0.37\textwidth,width=0.48\textwidth]{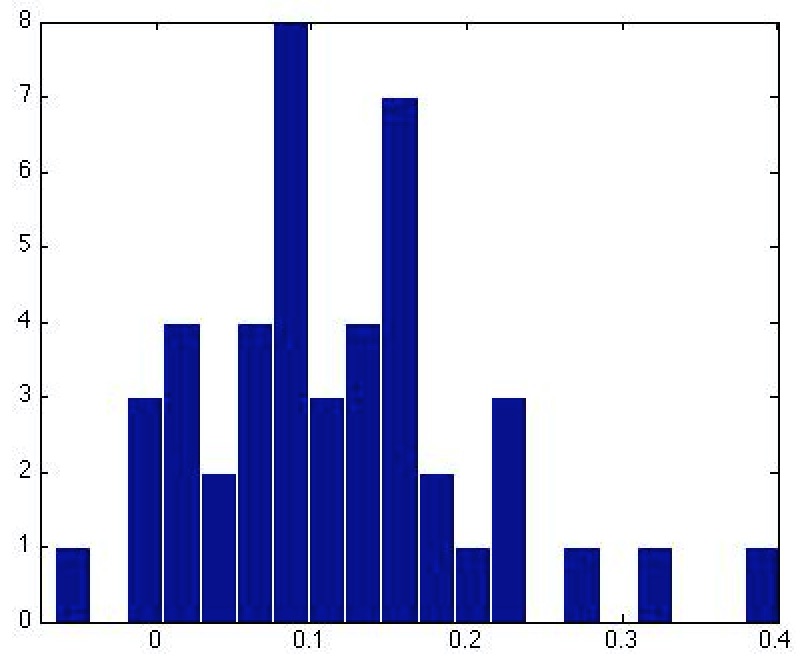}}	
        \caption{Binary classification accuracy. a) Comparison between i-vector (in the lower left triangle) and DCAR (in the upper right triangle) for each pairwise classification, with darker color indicating higher accuracy. b) Histogram of the accuracy improvement obtained by DCAR relative to i-vector across the 45 classifications.} \label{Fig:BC45AccIvDGMM} 
\end{figure*}

\section{Experimental Results} \label{SecExp}
\label{sec:experiments}

We evaluated the four representations under study in a combination of binary detection and multi-event detection tasks, described below. 

\subsection{Binary-Event Detection}
\label{sec:binary}

In the first experiment, we built 45 binary classifiers (one for each pair of events). We had two goals in conducting this experiment. The first was to compare two representation strategies: modeling GMMs on all training tracks, in this case as the first phase of the i-vector approach, vs.\ modeling GMMs on each training track, using DCAR. The second was to investigate \textit{how} the events are distinguished. 

As the graphs in Figure \ref{Fig:BC45AccIvDGMM} show, DCAR outperforms i-vector on most tasks.
On average, DCAR achieves an accuracy improvement of $10.74\%$ over i-vector (0.8293 vs.\ 0.7489) across the binary detection tasks. 
The win-tie-loss value for pairwise tests at the 0.05 significance level for DCAR against i-vector is 35-7-3; the win-tie-loss value at the 0.01 significance level is 40-2-3.

From these results, we can see that there are some event pairs that are particularly difficult to distinguish, such as Ev106--Ev107 and Ev107--Ev108. Considering the nature of the events involved, it could be argued that distinguishing between events with a \textit{Person Grooming an Animal} (Ev106), a \textit{Person Hand-Feeding an Animal} (Ev107), and a \textit{Person Landing a Fish} (Ev108) 
could be non-trivial even for humans. 
Nonetheless, compared with i-vector, our proposed DCAR increases binary classification accuracy even on these difficult pairs. 
This result demonstrates that modeling each audio file via a Gaussian mixture model is more suitable to characterizing audio content, and that integrating label information and local structure is useful in generating discriminative representations.

\subsection{Easy vs.\ Hard Cases}
\label{sec:easyhard}

To further explore how our proposed DCAR method performs on audio-based event detection under different difficulty levels,
we extracted two subsets from YLI-MED. Subset EC5 (``EasyCase'') contains five events (Ev101, Ev104, Ev105, Ev108, and Ev109) that are generally easy to distinguish from (most of) the others.
Subset HC4 (``HardCase'') contains four events (Ev103, Ev106, Ev107, and Ev110) that are more difficult to distinguish.\footnote{The division was made based on results from the experiments described in Subsections \ref{sec:binary} and \ref{sec:tendetect}, as well as \textit{a priori} understanding of the events' similarity from a human point of view. Because multiple criteria were used, Ev102 did not fall clearly into either category.}

\subsubsection{Dimensionality Tuning}

\begin{figure*}[!htb]
   \centering
   \subfigure[EC5]{
   \includegraphics[height=0.3\textwidth,width=0.47\textwidth]{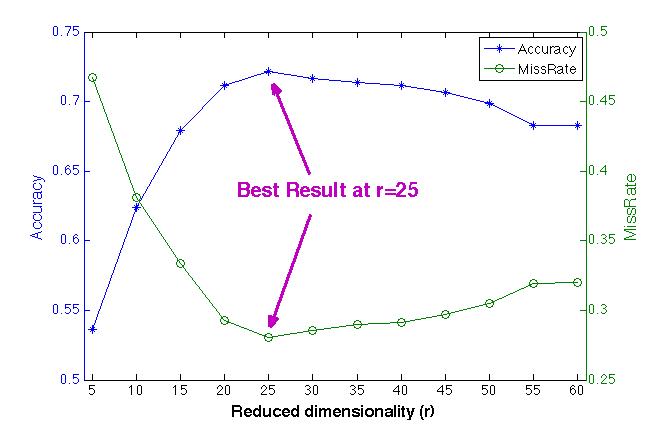}}	
   \subfigure[HC4]{
	\includegraphics[height=0.3\textwidth,width=0.47\textwidth]{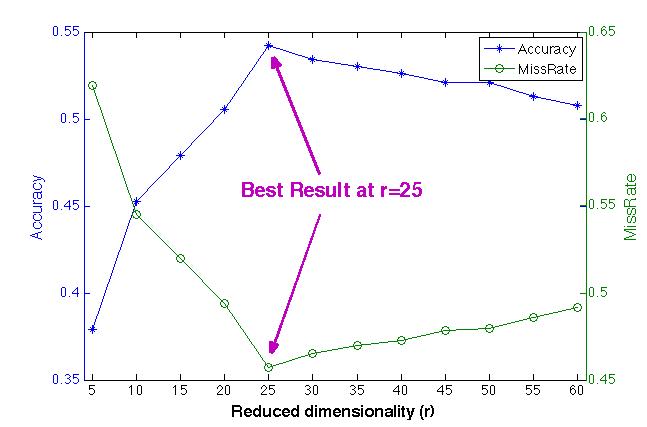}}	
    \caption{Effects of varying the parameter ($r$) in DCAR for the EasyCase subset (five-event detection), and for the HardCase subset (four-event detection) in terms of Accuracy and MissRate.} \label{Fig:VarDC5OC4VarR}
\end{figure*}

Before comparing DCAR with other representations, we conducted a set of multi-event detection experiments to study how the dimensionality parameter ($r$) affects DCAR under these two difficulty levels. Here, we used five-fold cross-validation on the training data to tune $r$.
The parameter was tuned from 5 to 60 with a step of 5. Combined $Accuracy$
and average $MissRate$ on EC5 and HC4 for each step are given in Figure \ref{Fig:VarDC5OC4VarR}. 

The results show that DCAR performs better as $r$ increases, reaches the best value at $r=25$ for both cases, and then again decreases in performance as $r$ grows larger.
We believe this is because
a smaller $r$ cannot optimally characterize the hidden structure of the data, while larger a $r$ may separate the structure 
into more dimensions until it is essentially equivalent to the original data, thus decreasing the efficacy of the representation.

\subsubsection{Easy and Hard Results}

Moving on to comparing DCAR with other state-of-the-art representations at these two difficulty levels, Table \ref{DC5OC4Compare4Ms} shows the multi-event detection performance of DCAR and three baseline representations---base GMM, mv-vector, and i-vector---in terms of $FScore$, $Accuracy$, $MissRate$ and $FAR$ on EC5 and HC4. 

\begin{table*}[!thb]
\centering
\caption{Comparison of four representations (mv-vector, i-vector, GMM, and DCAR) on multi-event detection for easy-to-distinguish events (EC5) and hard-to-distinguish events (HC4). (Best results in boldface.)}\label{DC5OC4Compare4Ms}
\begin{tabular}{l|cccc||cccc}
\hline
     & \multicolumn{4}{c||}{Subset EC5}  & \multicolumn{4}{c}{Subset HC4} \\ 
  Evaluation    & \multicolumn{4}{c||}{(Ev101,Ev104, Ev105, Ev108, Ev109)}  & \multicolumn{4}{c}{(Ev103,Ev106, Ev107, Ev110)} \\ \cline{2-9}
 Metric    &mv-vector &i-vector &GMM &DCAR   &mv-vector &i-vector &GMM &DCAR \\\hline
FScore($\uparrow$) &0.4773        &0.6415	&0.6670	&\bf{0.7067}   &0.4278	&0.2795	&0.4821	&\bf{0.5282} \\Accuracy($\uparrow$) &0.5455	&0.6828	&0.7131	&\bf{0.7434}  &0.4573	&0.2863	&0.5000	&\bf{0.5684}\\MissRate($\downarrow$) &0.5168	&0.3367	&0.3252	&\bf{0.2779} &0.5393	&0.6975	&0.4840	&\bf{0.4577}\\FAR($\downarrow$) &0.1136	&0.0785	&0.0730	&\bf{0.0647} &0.1788	&0.2409	&0.1684	&\bf{0.1496}\\
\hline
\end{tabular}
\end{table*}

For both subsets, DCAR consistently achieves the best results (marked in bold) on each evaluation metric, in comparison with the three baselines. (For $Accuracy$, p = 0.01 or better for pairwise comparisons of DCAR vs.\ mv-vector and i-vector, and p = 0.05 or better for DCAR vs.\ the base GMM, for both subsets. Significance is assessed using McNemar's two-tailed test for correlated proportions.)
However, interestingly, i-vector performs better than mv-vector on the EC5 data, but worse than mv-vector on the HC4 data (p = 0.0001 or better for $Accuracy$ comparisons for both subsets).

\subsubsection{DCAR, Variance, and Structure}

We can make a number of observations about these results. First, it seems that modeling GMM components for each audio track (as in the GMM, DCAR, and mv-vector representations\footnote{For these purposes, we can treat the mean and variance within the mv-vector as a GMM with one component.}) is more effective than modeling a GMM on all the training audio tracks together (as in i-vector) when the events are semantically related to each other (as in HC4).
We believe this is due to the fact that, in real-world applications (e.g., with user-generated content), each audio track may have a large variance. The set of strategies that model each track via GMM capture the hidden structure within each audio track, while the i-vector strategy may smooth away that structure (even between events), leading to a less useful representation.

Second, GMM and DCAR perform better than mv-vector on both subsets. 
We believe this indicates that one mixture component (as in mv-vector) may not sufficiently capture the full structure of the audio; in addition,
vectorizing the mean and variance inevitably distorts the intrinsic geometrical structure among the data. 
Third, DCAR outperforms the base GMM. As we described in Section \ref{SecDGMMwhole}, DCAR begins by extracting such a GMM model, but it also takes into account the label information and the intrinsic nearest neighbor structure among the audio files when modeling the training data, and outputs a mapping function to effectively represent the test data. 

In sum, these experimental results further confirm that discriminative dimensionality reduction is beneficial for characterizing the distinguishing information for each audio file, leading to a better representation.

\subsection{Ten-Event Detection}
\label{sec:tendetect}

Because event detection is a kind of supervised learning,
learning becomes more difficult as the number of events increases.
In the next experiment, we again compared the proposed DCAR model with the three baseline representations, this time on a ten-event detection task. As before, the parameters for each method were tuned by cross-validation on the training data. 
Table \ref{C10perECompare4Ms} gives the detection performance on each event and their average over the ten events in terms of $FScore$ and $MissRate$.

\begin{table*}[!thb]
\centering
\caption{Per-event comparison of detection performance (as FScore and MissRate) using four representations: mv-vector, i-vector, GMM, and DCAR. (Best results in boldface; second-best underlined.)}\label{C10perECompare4Ms}
\begin{tabular}{c|cccc||cccc}
\hline
            & \multicolumn{4}{c||}{FScore ($\uparrow$)}     & \multicolumn{4}{c}{MissRate ($\downarrow$)}    \\\cline{2-9}
               &mv-vector &i-vector &GMM &DCAR       &mv-vector &i-vector &GMM &DCAR \\\hline 
Ev101 &0.7259	&\bf{0.7842}	&0.7303	&\underline{0.7835}	&0.2824	&0.1679	&\underline{0.1527}	&\bf{0.1298} \\
Ev102 &0.2837	&0.3396	&\underline{0.3651}	&\bf{0.4603}	&0.5918	&0.6327	&\underline{0.5306}	&\bf{0.4082}\\
Ev103 &0.2178	&\underline{0.2569}	&0.2410	&\bf{0.3820}	&0.7179	&\underline{0.6410}	&0.7436	&\bf{0.5641}\\
Ev104 &0.4274	&\underline{0.6206}	&0.6000	&\bf{0.6207}	&0.6063	&0.4331	&\underline{0.3622}	&\bf{0.3621}\\
Ev105 &0.3354	&0.3899	&\bf{0.5714}	&\underline{0.5178}	&0.6932	&0.6477	&\bf{0.3864}	&\underline{0.4205}\\
Ev106 &0.1964	&0.1835	&\underline{0.2963}	&\bf{0.3750}	&0.7105	&0.7368	&\underline{0.6842}	&\bf{0.6053}\\
Ev107 &\underline{0.3850}	&0.3298	&0.3250	&\bf{0.4024}	&\bf{0.6814}	&0.7257	&0.7699	&\underline{0.7080}\\
Ev108 &0.3191	&0.3853	&\underline{0.3878}	&\bf{0.4231}	&0.6341	&\underline{0.4878}	&0.5366	&\bf{0.4634}\\
Ev109 &0.4211	&\underline{0.5028}	&0.4286	&\bf{0.5176}	&0.6667	&\bf{0.5833}	&0.6667	&\underline{0.5926}\\
Ev110 &0.0833	&\underline{0.2299}	&\bf{0.2857}	&0.2162	&0.9091	&\underline{0.7727}	&\bf{0.7500}	&0.8182\\ \hline
Average   &0.3395	&0.4023	&0.4231	&\bf{0.4699} &0.6494	&0.5829	&0.5583	&\bf{0.5072}\\
\hline
\end{tabular}
\end{table*}

For all of the individual events and on average, DCAR achieves superior or competitive performance. 
DCAR also performs better in terms of $Accuracy$ and $FAR$. The overall $Accuracy$ scores for mv-vector, i-vector, GMM, and DCAR are 0.3907, 0.4640, 0.4923, and {\bf{0.5321}}, respectively (p = 0.01 for DCAR vs.\ each baseline [McNemar's two-tailed]), and the average $FAR$ scores are 0.0674, 0.0593, 0.0570, and {\bf{0.0523}}. 

Although other representations may perform as well or better on some particular events,
DCAR consistently outperforms the other representations for all evaluation metrics on the \textit{average} or \textit{overall} scores (an average of more than 8\%  gain on all metrics relative to the second best representation).  These results further demonstrate that modeling each audio file via GMM and then integrating both label information and local structure are beneficial to constructing a discriminative audio representation for event detection.

In addition to comparing results across the four methods for ten-event detection, we also experimented with applying feature reduction methods at the frame level before training the GMM, using PCA \cite{Ji05} and linear discriminant analysis (LDA) \cite{McLachlanG2004}.
(As an alternative to DCAR's approach to dimensionality reduction.) The number of principal components ($r$) in PCA was tuned from 5 to 60 with a step of 5. For LDA, $r=L-1$, where $L$ is the number of events. Average or overall results are given in Table \ref{tab:redcomp}. 
The results with PCA are a little better than GMM without PCA, but the accuracy difference is not statistically significant (p = 0.7117, McNemar's two-tailed). Results with LDA are much worse than GMM without LDA. We hypothesize that the main reason for the poor performance of LDA+GMM is that LDA only considers $L-1$ components, which is usually too few to capture sufficient information for later GMM training.

\begin{table*}[!thb]
\centering
\caption{Comparison of detection performance for GMM representations with and without pre-training feature reduction.}
\label{tab:redcomp}
\begin{tabular}{c|ccc}
\hline
Evaluation & GMM & PCA+GMM & LDA+GMM \\ 
Metric &  & $r=30$ & $r=9$ \\ \hline
FScore ($\uparrow$) & 0.4231 & 0.4293 & 0.3278 \\
Accuracy ($\uparrow$) & 0.4923 & 0.4987 & 0.3419 \\
MissRate ($\downarrow$) & 0.5583 &  0.5508 & 0.6574 \\
FAR ($\downarrow$) & 0.0570 & 0.0562 & 0.0935 \\ \hline
\end{tabular}
\end{table*}

\subsection{Intra-Event Variation}

Delving deeper into how the effectiveness of a representation may depend on variable characteristics of audio tracks, 
we looked at the degree to which some test tracks in YLI-MED could be classified more accurately than others. 

\subsubsection{Variable Performance Within Events}
We took the predicted result for each test audio track in the experiments with four representations described in Subsection \ref{sec:tendetect} and calculated how many of the representations made correct predictions for that track. 

\begin{figure}[!htb]
   \centering
 \includegraphics[width=0.7\textwidth]{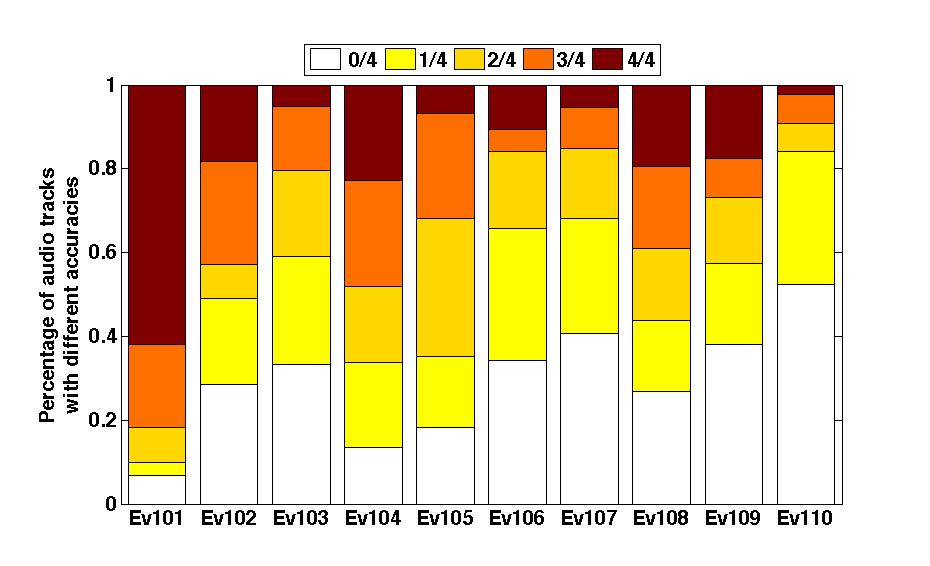} 
 \caption{Per-event percentages of test tracks that are correctly classified by how many representations.
    \textit{Acc} is the proportion of representation types that correctly classify a given audio track. 
     } \label{Fig:PUC4Ms}
\end{figure}

Figure \ref{Fig:PUC4Ms} shows the distribution of the number of representations making accurate predictions for each audio track, broken down by event.
Generally, there is wide variation in accuracy among audio files belonging to the same event, with the exception of Ev101 (Birthday Party); this suggests that Ev101 may have distinctive audio characteristics that lead to more consistent classification. It is worth noting that there are some audio files that are never correctly classified by any of the representations (i.e., where \textit{acc} = 0). For example, more than 50\% of the audio tracks for Ev110 (Working on a Woodworking Project)  could not be correctly classified by any of the four representations. This situation highlights a challenging property of the event detection task: some of the events are quite confusable due to their inherent characteristics.
For example, Ev103 (Getting a Vehicle Unstuck) may have similar audio properties to Ev110 (Working on a Woodworking Project) in that both are likely to involve sounds generated by motors. This is also the
reason we included Ev103 and Ev110 in the ``Hard Case'' HC4 dataset for the experiments described in Section \ref{sec:easyhard}. 

\subsubsection{Relationship to Annotator Confidence}

When the YLI-MED videos were collected, three annotators 
were asked to give a confidence score for each video, chosen from \{1,2,3\}, with 1 being ``Not sure'' and 3 being ``Absolutely sure'' that the video contains an example of the event in question. The average of the three scores can be used as an indicator of how easily classifiable a given video is with respect to the event category, from the perspective of human beings.

\begin{figure}[!htb]
   \centering
    \includegraphics[width=0.7\textwidth]{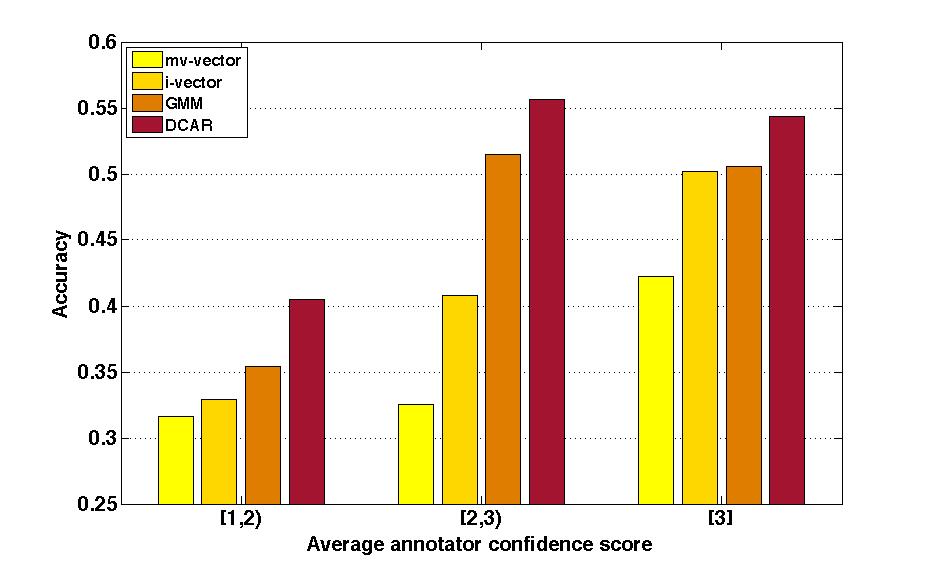} 
    \caption{Accuracy of the four representations for audio from videos with varying annotator
    confidence.} \label{Fig:AccConfidence4Ms}
\end{figure}

Figure \ref{Fig:AccConfidence4Ms} shows the combined $Accuracy$ scores of different representations for audio tracks from videos in three confidence ranges (79 of the test audio tracks in the range [1,2), 169 in the range [2,3), and 530 with a average confidence of [3]). DCAR achieves the best performance in every range. However, it is interesting that i-vector shows a notable improvement with each increment of increasing annotator confidence (+24.1\% between [1,2) and [2,3) and +22.9\% between [2,3) and [3]), while DCAR shows a dramatic improvement between low- and intermediate-confidence videos but performs similarly on intermediate- and high-confidence videos (+37.3\% for the first step, but -2.3\% for the second); GMM follows the same pattern (+45.6\% then -1.8\%). The mv-vector approach shows yet a different pattern, performing similarly poorly on all but the high-confidence videos (only +2.8\% improvement for the first step, but +29.8\% for the second).

These differing patterns may indicate that the i-vector approach  
is more sensitive to particular audio cues associated with the characteristics of an event that humans find most important in categorizing it, while a lower threshold of cue distinctiveness overall is required for GMM and DCAR.
On the other hand, the results in Sections \ref{sec:easyhard} and \ref{sec:tendetect} suggest that modeling audio files with only one mixture component, as in mv-vector, generally cannot sufficiently capture the full structure of the audio signal, 
but it may be that it can capture that structure at least somewhat more often when the signal is more distinctive (i.e., it has a particularly high distinctiveness threshold).   

In other words, in cases where humans do not consider the events occurring in videos to be clear or prototypical examples of the target category, and thus those videos are less likely to have plentiful audio cues distinctive to that event category, a more 
discriminative representation may be required to improve event detection performance.\footnote{The experiments here did not include any negative test tracks; the task was simply to 
identify which of the target events each video's contents are \textit{most} like.
If negative examples were included, it might be less clear what constitutes the ``best'' performance in categorizing videos that are on the borders of their categories.} 
Fortunately, it appears that the proposed DCAR method can partially address this problem, as shown by its relatively high performance on lower-confidence videos.

\section{Conclusions and Future Work}
\label{sec:conc}

In this article, we have presented a new audio representation, DCAR, and demonstrated its use in event detection.
One distinguishing characteristic of the DCAR method is that it can capture the variability within each audio file. Another is that it achieves better discriminative ability by integrating label information and the graph of the components' nearest neighbors among the audio files, i.e., it can successfully characterize both global and local structure among audio files.
Representing audio using the proposed DCAR notably 
improves performance on event detection, as compared with state-of-the-art representations (an average of more than 8\% relative gain for ten-event detection and more than 10\% gain for binary classification across all metrics).

The proposed representation benefits from leveraging global and local structure within audio data; however, videos are of course \textit{multi}modal. Other data sources such as visual content, captions, and other metadata can provide valuable information for event detection; we therefore plan to extend the current model by incorporating such information. Within audio, we also hope to evaluate the use of DCAR for other related tasks, such as audio scene classification (for example, testing it with the DCASE acoustic scenes dataset \cite{SdGdBeLmPm15}). 

Related work in audio (e.g., Barchiesi et al.\ 2015 \cite{BdGdSdPm15}) has demonstrated that the temporal evolution of different events plays an important role in audio analysis; another possible direction for expanding DCAR is to take into consideration complex temporal information in modeling video events. Last but not least, we might explore extracting the information-rich  
segments from each audio track rather than modeling the whole track.

\section*{Acknowledgments}
This work was partially supported by the NSFC (61370129, 61375062), the PCSIRT (Grant IRT201206), and a collaborative Laboratory Directed Research \& Development grant led by Lawrence Livermore National Laboratory (U.S.\ Dept.\ of Energy contract DE-AC52-07NA27344). Any findings and conclusions are those of the authors, and do not necessarily reflect the views of the funders.

\bibliographystyle{abbrv}
\bibliography{main}

\appendix

\section*{Appendices}

\section{Comparison of LEM, AIRM, and Stein Divergence}
\label{app:comparison}

To empirically determine which metric (LEM, AIRM, or Stein Divergence) is  
most appropriate for measuring the distance between GMM covariance matrices, we calculated what percentage of each component's $k$-nearest neighbors belong to the same event as the that component as in \ref{PCNN}.
\begin{equation}\label{PCNN}
PC(k) = \frac{1}{N}\sum_{i=1}^N \frac{|Y|Y\in N_k(X_i) \bigcap| \ell(Y)=\ell(X_i)|}{k}
\end{equation}
$N$ is the number of training mixture components.
$N_k(X_i)$ gives the $k$-nearest neighbor set for component $X_i$ obtained 
by the given metric.
$\ell(Y)=\ell(X_i)$ indicates that $Y$ and $X_i$ belong to the same event. 
A high value for $PC$ indicates that the metric is suitable for characterizing the structure of the components, 
in the sense that the result is similar to the external structure given by the label information. 

\begin{figure}[!htb] 
   \centering
   \includegraphics[width=0.6\textwidth]{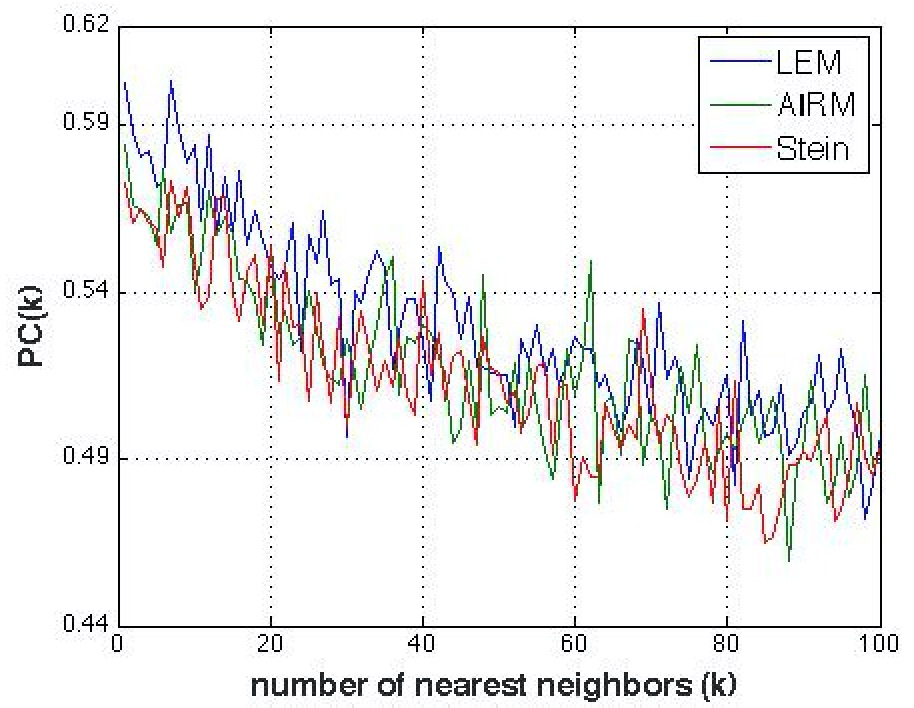}
    \caption{Comparison of LEM, AIRM, and Stein Divergence on the covariance matrices of audio GMM components (extracted from YLI-MED training data). The x-axis indicates the number of nearest neighbors ($k$), and the y-axis indicates the average percentage of neighbors belonging to the same event as the target component.} \label{Fig:Metric}
\end{figure}

Figure \ref{Fig:Metric} plots the values for $PC(k)$ obtained by the three metrics when varying the target number of neighbors ($k$).
The plot shows  
that, overall, LEM
achieves higher values for $PC$ than AIRM or Stein Divergence,i.e., that the nearest neighbors selected according to LEM more often fall into the same event than nearest neighbors selected according to either AIRM or Stein Divergence.

Therefore, we use LEM to compute the distance between the covariance matrices according to the formula in (\ref{DisSigma}) above, repeated here as (\ref{DisSigmaAgain}): 
\begin{equation}\label{DisSigmaAgain}
\delta_{\Sigma}^2(\Sigma_i,\Sigma_j)=\| \log(\Sigma_i)-\log(\Sigma_j)\|_F^2
\end{equation}

\section{Proof: The Rotational Invariance of $\mathbf{F}$}
\label{app:invariance}

Given the objective function $\mathbf{F}(\mathbf{W})$ in (\ref{DLDGMM2}), repeated here as (\ref{DLDGMM2Again}), 
and any rotation matrix $\mathbf{R}\in SO(r)$ (i.e., $\mathbf{R}\mathbf{R}^T=\mathbf{R}^T\mathbf{R}=\mathbf{I}_r$), we can show that $\mathbf{F}(\mathbf{W})=\mathbf{F}(\mathbf{WR})$. 
\begin{equation}\label{DLDGMM2Again}
\begin{split}
\mathbf{F}(\mathbf{W}) = 
& \min\limits_{\mathbf{W}^T\mathbf{W}=I_{r}}\sum_{i,j}A_{ij}\Big (\lambda \|\mathbf{W}^T(\mu_i-\mu_j)\|_F^2 \\
& + \|\log(\mathbf{W}^T\Sigma_i\mathbf{W}) - \log(\mathbf{W}^T\Sigma_j\mathbf{W})\|_F^2\Big )
\end{split}
\end{equation}

\begin{proof}
\label{pr:rotinvariance}
According to the definition of $\mathbf{F}(\mathbf{W})$ in (\ref{DLDGMM2Again}), we can write $\mathbf{F}(\mathbf{WR})$ as
\begin{equation}\label{DLDGMM3}
\begin{split}
\mathbf{F}(\mathbf{WR}) = 
& \min\limits_{\mathbf{W}^T\mathbf{W}=I_{r}}\sum_{i,j}A_{ij}\Big (\lambda \|(\mathbf{WR})^T(\mu_i-\mu_j)\|_F^2 \\ \nonumber
& + \|\log((\mathbf{WR})^T\Sigma_i\mathbf{WR}) - \log((\mathbf{WR})^T\Sigma_j\mathbf{WR})\|_F^2\Big )
\end{split}
\end{equation}
We set
\begin{equation}
\begin{split}
\mathbf{F1}_{ij}(\mathbf{WR}) &= \|(\mathbf{WR})^T(\mu_i-\mu_j)\|_F^2\\
&=Tr\Big [(\mu_i-\mu_j)^T (\mathbf{WR}) (\mathbf{WR})^T (\mu_i-\mu_j) \Big ]\\
& = Tr\Big [(\mu_i-\mu_j)^T \mathbf{WR}\mathbf{R}^T\mathbf{W}^T(\mu_i-\mu_j) \Big ]\\
& =  Tr\Big [(\mu_i-\mu_j)^T \mathbf{W}\mathbf{W}^T(\mu_i-\mu_j) \Big ]\\
& =  \|(\mathbf{W})^T(\mu_i-\mu_j)\|_F^2\\
& = \mathbf{F1}_{ij}(\mathbf{W})
\end{split}
\end{equation}
Since  $0\prec \mathbf{W}^T\Sigma_i\mathbf{W} \in \mathcal{S}ym_{r}^+$ is symmetric positive-definite, and the log-euclidean Metric has the properties of Lie group bi-invariance and similarity invariance), i.e., 
$$\|\log(X)-\log(Y)\|_F^2 = \|\log(\mathbf{R}^TX\mathbf{R})-\log(\mathbf{R}^TY\mathbf{R})\|_F^2$$ 
then
\begin{equation}
\begin{split}
\mathbf{F2}_{ij}(\mathbf{WR}) & =\|\log((\mathbf{WR})^T\Sigma_i\mathbf{WR}) - \log((\mathbf{WR})^T\Sigma_j\mathbf{WR})\|_F^2\\
& =\|\log(\mathbf{R}^T(\mathbf{W}^T\Sigma_i\mathbf{W})\mathbf{R}) - \log(\mathbf{R}^T(\mathbf{W}^T\Sigma_j\mathbf{W})\mathbf{R})\|_F^2\\
&=\|\log((\mathbf{W}^T\Sigma_i\mathbf{W}) - \log(\mathbf{W}^T\Sigma_j\mathbf{W})\|_F^2\\
& = \mathbf{F2}_{ij}(\mathbf{W})
\end{split}
\end{equation}
Thus
\begin{equation}
\begin{split}
\mathbf{F}(\mathbf{WR}) &=\min\limits_{\mathbf{W}^T\mathbf{W}=I_{r}}\sum_{i,j}A_{ij}\big (\lambda \mathbf{F1}_{ij}(\mathbf{WR})  + \mathbf{F2}_{ij}(\mathbf{WR})  \big )\\
&=\min\limits_{\mathbf{W}^T\mathbf{W}=I_{r}}\sum_{i,j}A_{ij}\big (\lambda \mathbf{F1}_{ij}(\mathbf{W})  + \mathbf{F2}_{ij}(\mathbf{W})  \big )\\
&=\mathbf{F}(\mathbf{W}) 
\end{split}
\end{equation}
as claimed.
\end{proof}

\section{Computing the Derivative of $\bm F2$}
\label{app:F2}

\begin{equation*}
	\begin{split}
	\nabla_{\mathbf{W}}\mathbf{F2}_{ij}
	&=\nabla_{\mathbf{W}} \Big (2Tr\big (\log(\mathbf{W}^T\Sigma_i\mathbf{W})-\log(\mathbf{W}^T\Sigma_j\mathbf{W})\big ) \Big) \\
	& \quad \times \Big (\log(\mathbf{W}^T\Sigma_i\mathbf{W})-\log(\mathbf{W}^T\Sigma_j\mathbf{W})\Big)
	 \\&=\nabla_{\mathbf{W}} \Big(2\log det(\mathbf{W}^T\Sigma_i\mathbf{W})-2\log det(\mathbf{W}^T\Sigma_j\mathbf{W})\Big )
	 \\& \quad \times \Big(\log(\mathbf{W}^T\Sigma_i\mathbf{W})-\log(\mathbf{W}^T\Sigma_j\mathbf{W})\Big)
	\\&=2\Big ( \big (\nabla_{\mathbf{W}} Tr(\mathbf{W}^T\Sigma_i\mathbf{W}) \big )(\mathbf{W}^T\Sigma_i\mathbf{W})^{-1} 
	\\& \quad \quad  - \big (\nabla_{\mathbf{W}} Tr(\mathbf{W}^T\Sigma_j\mathbf{W}) \big )(\mathbf{W}^T\Sigma_j\mathbf{W})^{-1}\Big )
	\\& \quad \times  \Big(\log(\mathbf{W}^T\Sigma_i\mathbf{W})-\log(\mathbf{W}^T\Sigma_j\mathbf{W})\Big)
	\\&=4\Big (\Sigma_i\mathbf{W}(\mathbf{W}^T\Sigma_i\mathbf{W})^{-1}-\Sigma_j\mathbf{W}(\mathbf{W}^T\Sigma_j\mathbf{W})^{-1}\Big ) 
	\\& \quad \times \Big(\log(\mathbf{W}^T\Sigma_i\mathbf{W})-\log(\mathbf{W}^T\Sigma_j\mathbf{W})\Big ).
	\end{split}
\end{equation*}

\end{document}